\documentclass[aps,prl,reprint,showpacs,floatfix]{revtex4-1}

\usepackage[cyr]{aeguill}
\usepackage[T1]{fontenc}
\usepackage[latin1]{inputenc}
\usepackage{amsmath}
\usepackage{amssymb}
\usepackage{graphicx}

\usepackage[svgnames]{pstricks}
\usepackage{pst-3dplot,pst-grad}

\begin{document}

\title{Tricritical point and wing structure in the itinerant ferromagnet UGe$_2$}
\author{V.~Taufour}
\author{D.~Aoki}%
	\email{dai.aoki@cea.fr}
\author{G.~Knebel}
\author{J.~Flouquet}

\affiliation{INAC/SPSMS, CEA-Grenoble, 17 rue des Martyrs, 38054 Grenoble, France}%

\date{\today}

\begin{abstract}
Precise resistivity measurements on the ferromagnetic superconductor UGe$_2$ under pressure $p$ and magnetic field $H$ reveal a previously unobserved change of the anomaly at the Curie temperature. Therefore, the tricritical point (TCP) where the paramagnetic-to-ferromagnetic transition changes from a second order to a first order transition is located in the $p$-$T$ phase diagram. Moreover, the evolution of the TCP can be followed under the magnetic field in the same way. It is the first report of the boundary of the first order plane which appears in the $p$-$T$-$H$~phase diagram of weak itinerant ferromagnets. This line of critical points starts from the TCP and will terminate at a quantum critical point. These measurements provide the first estimation of the location of the quantum critical point in the $p$-$H$ plane and will inspire similar studies of the other weak itinerant ferromagnets.
\end{abstract}

\pacs{PACS}
\maketitle

Quantum critical points (QCP) emerge when a second order phase transition occurs at zero temperature. They have been studied intensively~\cite{Hertz1976,Millis1993,Lohneysen2007}, due to fascinating phenomena, such as unconventional superconductivity and non-Fermi liquid behavior, which are expected close to the QCP. One famous example of a second order phase transition is the paramagnetic-ferromagnetic (PM-FM) transition and intensive efforts have been made to drive such a transition to $0$~K in order to study the quantum criticality. Surprisingly, approaching $0$~K by chemical substitution and/or pressure, the PM-FM transition becomes first order in all materials studied so far~: UGe$_2$~\cite{Pfleiderer2002}, ZrZn$_2$~\cite{Uhlarz2004}, CoS$_2$~\cite{Goto1997}, or SrRuO$_3$~\cite{Uemura2007}. This apparently generic result~\cite{Belitz2007} is in contrast with the theoretical prediction by Hertz~\cite{Hertz1976} that the quantum ferromagnetic transition in metals should be of second order. Metamagnetism is often observed in the paramagnetic regime close to the ferromagnetic instability~\cite{Huxley2000,Pfleiderer2002,Uhlarz2004,Goto1997,Thessieu1995}.

The typical phase diagram of these compounds has a tricritical point (TCP) where the transition changes from second to first order and the first order surface at $H=0$ bifurcates at high pressure into two first order surfaces or ``wings''~\cite{Griffiths1973}. These first order surfaces are limited by a second order transition line that goes to $T=0$~K at a quantum critical end point (QCEP). A QCEP differs from a QCP by the absence of spontaneous symmetry breaking~\cite{Grigera2001}. To date, the wings have been drawn in the $p$-$T$-$H$~space theoretically~\cite{Belitz2005}. Thus, the observed metamagnetic behaviors above the TCP can be identified either to the first order surfaces or to the associated crossover above the second order critical line.
Such a three-dimensional phase diagram is often presented only in a qualitative form~\cite{Uhlarz2004,Kimura2004,Pfleiderer2001,Belitz2005,Yamada2007,Rowley2010}. One difficulty for theoretical studies is to get a quantitative prediction; this arises from the fact that the parameters are complicated functions of the pressure or chemical doping~\cite{Belitz2005}. For experimental studies, a precise tuning of the three parameters is necessary. Moreover, the extension of the ``wings'' in the $p$-$T$-$H$~space can be very small~\cite{Uhlarz2004,Belitz2005}. In this paper, we report experimental observation of the wings and TCP coordinates in the ferromagnetic superconductor UGe$_2$ ($T_{\textrm{TCP}}\approx24$~K and $p_{\textrm{TCP}}\approx1.42$~GPa). The position of the QCEP in the $p$-$H$ plane is also estimated by extrapolation.

\begin{figure}[!ht]
\begin{center}
\includegraphics[width=6.3cm]{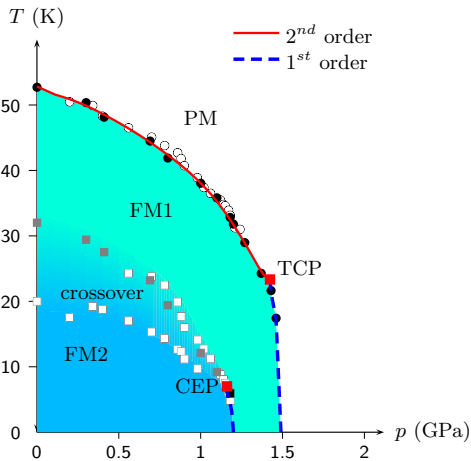}
\caption{\label{fig:diagpT}(Color online) Pressure-temperature phase diagram of UGe$_2$ drawn from our results of thermal expansion (open symbols) and resistivity (full symbols). The tricritical point TPC separate the second order to the first order paramagnetic-ferromagnetic PM-FM transition. The critical end point CEP separate the first order FM1-FM2 transition to the crossover regime. For clarity, superconductivity is not shown.}
\end{center}
\end{figure}

UGe$_2$ is one of the most extensively studied ferromagnetic superconductors. It crystallizes in an orthorhombic structure with the space group $Cmmm$. The ferromagnetic moment is directed along the a-axis, and the ordered moment at $T=2$~K is $M_0\approx1.4$~$\mu_B/$U at ambient pressure. The Curie temperature $T_{\textrm{C}}$ decreases with pressure and disappears at the critical pressure $p_c\approx1.49$~GPa [Fig.~\ref{fig:diagpT}]. From thermal expansion measurements, the transition is of second order at low pressure~\cite{Nishimura1994}, but near $p_c$ the transition becomes first order~\cite{Huxley2000,Pfleiderer2002}. The phase diagram is even more complex~: pressure  measurements at $T=2$~K indicates that the magnetization jumps at a critical pressure $p_x\approx1.19$~GPa from a low pressure FM2 phase, with the large moment of $M_0\approx1.4$~$\mu_B/$U, to a FM1 phase with $M_0\approx0.9$~$\mu_B/$U. This FM1-FM2 transition is of first order~\cite{Pfleiderer2002}. It is now well established that there is no phase transition between FM1 and FM2 at ambient pressure but a very broad crossover~\cite{Hardy2009}. Thus, the FM1-FM2 transition disappears at a critical end point (CEP) above which a crossover regime is observed ($T_{\textrm{CEP}}\approx7$~K and $p_{\textrm{CEP}}\approx1.16$~GPa~\cite{Taufour2010}).

Just above $p_c$, when the magnetic field is applied along the easy axis of the magnetization (a-axis), a metamagnetic transition from the PM to the FM1 state is observed at $H_c$~\cite{Huxley2000,Pfleiderer2002}. Further increasing $H$ leads to a recovery of the FM2 phase for $H>H_x$.

Superconductivity has been discovered only in the ferromagnetic state with the maximum $T_{\textrm{sc}_{\textrm{max}}}$ of the superconducting transition temperature $T_{\textrm{sc}}$ very close to $p_x$~\cite{Saxena2000}. It has been theoretically proposed that spin or charge density waves, which are favorable for the appearance of superconductivity, is developed at $p_x$~\cite{Watanabe2002}. However, there is no experimental evidence for such a hypothesis yet.

In the present study, several high-quality single crystals of UGe$_2$ were grown by the Czochralski method in a tetra-arc furnace. The samples were cut by a spark cutter and checked by x-ray Laue diffraction, resistivity, thermal expansion, and specific heat measurements. The residual resistivity ratio is higher than 300, indicating the high quality of the samples. Here we report only the results of the resistivity measurements under pressure and magnetic field. Pressure was applied via a NiCrAl-CuBe hybride piston-cylinder cell with Daphne oil 7373 as a pressure-transmitting medium. The pressure was determined by measuring the superconducting transition of Pb by AC susceptibility. Electrical resistivity was measured down to $2$~K and at fields up to $9$~T, employing the four probe AC method with current parallel to the a-axis. Magnetic field was applied along the a-axis, which corresponds to the magnetization easy axis.

Two types of anomalies at $T_{\textrm{C}}$ (a peak or a minimum in $d\rho/dT$) are observed. The different behavior of these anomalies under the magnetic field allow us to determine the location of the TCP ($T_{\textrm{TCP}}\approx24$~K and $p_{\textrm{TCP}}\approx1.42$~GPa), to draw the surfaces of first-order transitions and the line of critical points that forms a boundary of these surfaces.

\begin{figure}[!htb]
\begin{center}
\includegraphics[width=8.6cm]{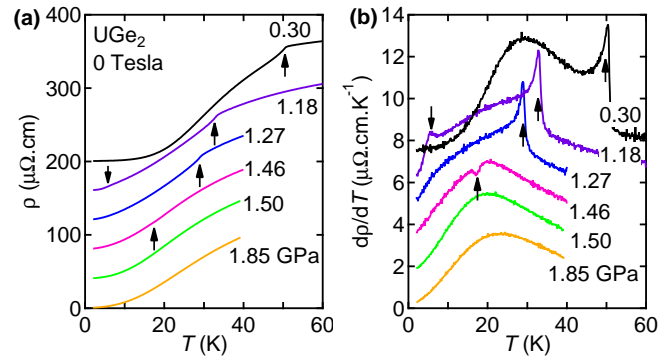}
\caption{\label{fig:rodrodT0Teslaab}(Color online) Temperature dependence of the electrical resistivity (a) and its temperature derivative (b) at different pressures. Down and up arrows indicate $T_x$ and $T_{\textrm{C}}$, respectively. The data are offset for clarity.}
\end{center}
\end{figure}

The temperature dependence of electrical resistivity at zero field is presented in Fig.~\ref{fig:rodrodT0Teslaab}(a). The different cases are clearly observed~: at $0.30$~GPa we detect only the Curie temperature $T_{\textrm{C}}$; at $1.18$~GPa, the transitions PM-FM1 at $T_{\textrm{C}}$ and FM1-FM2 at $T_x$; at $1.27$ and $1.46$~GPa, only the PM-FM1 transition; and above $p_c\approx1.49$~GPa, only the PM regime. The phenomena are more obvious taking into account the temperature derivative of the resistivity $d\rho/dT$ [Fig.~\ref{fig:rodrodT0Teslaab}(b)]. The most striking point is the change of the anomaly at $T_{\textrm{C}}$ going from a sharp positive maximum of $d\rho/dT$ at low pressure to a small negative minimum close to $p_c$.

The sharp positive maximum of $d\rho/dT$ indicates that the resistivity is abruptly suppressed below $T_{\textrm{C}}$. This is usually observed in ferromagnetic metals where the resistivity due to the spin disorder scattering is scaled by the bulk magnetization as $1-[M(T)/M(0)]^2$~\cite{Fisher1968}. In the other case, the negative peak of $d\rho/dT$ at $T_{\textrm{C}}$ is indicative of a hump of resistivity. The anomaly in the temperature dependence of $d\rho/dT$ changes drastically and indicates the switch at the TCP from a second order to a first order transition.

\begin{figure*}[!htb]
\begin{center}
\includegraphics[width=17.92cm]{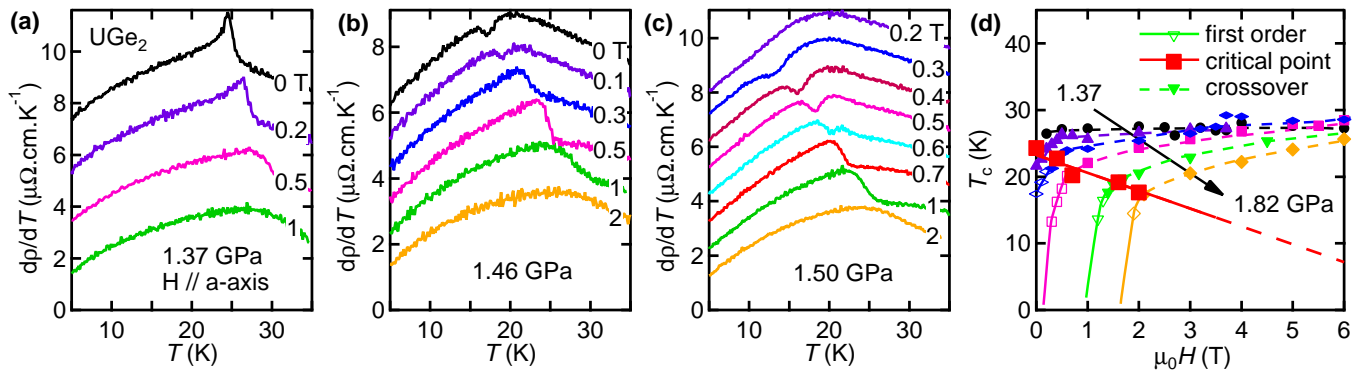}
\caption{\label{fig:drodTabcd}(Color online) Temperature derivative of resistivity $d\rho/dT$ for three typical pressures~: (a)~$p<p_{\textrm{TCP}}$ ; (b)~$p_{\textrm{TCP}}<p<p_c$, and (c)~$p>p_c$. The data are offset for clarity. See text for discussion. (d)~Magnetic field dependence of $T_{\textrm{C}}$ at different pressures ($1.37$, $1.43$, $1.46$, $1.50$, $1.65$, and $1.82$~GPa). $T_{\textrm{C}}$ is defined at the optimum of the positive or negative anomaly. The line of critical point $H_{\textrm{CP}}(T,p)$ separates the different regimes. The PM-FM1 transition will occur only below $H_{\textrm{CP}}$~: full lines, open symbols, negative anomalies. For $H>H_{\textrm{CP}}$ (crossover)~: dashed lines, full symbols, positive anomalies. The $H_{\textrm{CP}}(T,p)$ line can be extrapolated to $0$~K at the QCEP.}
\end{center}
\end{figure*}

We will now focus on the field dependence of the resistivity anomaly. Three different cases are presented in Fig.~\ref{fig:drodTabcd}~: $p<p_{\textrm{TCP}}$, $p_{\textrm{TCP}}<p<p_c$, and $p_c<p$.

In Fig.~\ref{fig:drodTabcd}(a), $p<p_{\textrm{TCP}}$, the second order PM-FM1 transition is observed as a positive peak in $d\rho/dT$. An applied magnetic field smears the anomaly out, since the applied magnetic field itself breaks the time reversal symmetry, and the peak of $d\rho/dT$ at $T_{\textrm{C}}$ is quickly broadened. As usual for conventional metallic ferromagnets, $T_{\textrm{C}}$ determined by the maximum of $d\rho/dT$ slightly increases under magnetic field [Fig.~\ref{fig:drodTabcd}(d)].

In contrast, Fig.~\ref{fig:drodTabcd}(b) shows $p_{\textrm{TCP}}<p<p_{c}$. At zero field the minimum in $d\rho/dT$ indicates the first order transition. But increasing $H$ to $H_{\textrm{CP}}\approx0.3$~T leads to the recovery of the second order like anomaly which smears out for higher fields.

Above $p_{c}\approx1.49$~GPa, no anomaly is detected for $H<H_c=0.2$~T at $1.50$~GPa (paramagnetic state); see Fig.~\ref{fig:drodTabcd}(c). With applied magnetic field above $H_c$, a minimum appears (first order PM-FM1 transition), but above $H_{\textrm{CP}}\approx0.7$~T, this negative anomaly suddenly becomes positive, indicating the change from first to second order transition.

\begin{figure}[!bht]
\begin{center}
\includegraphics[width=5.5cm]{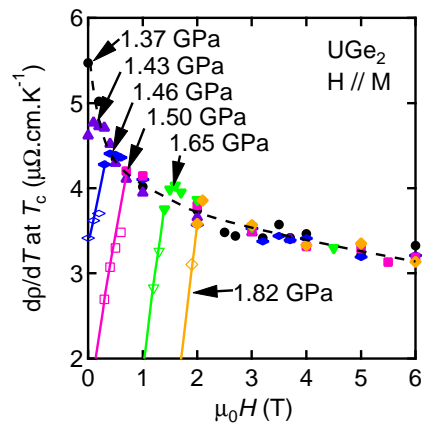}
\caption{\label{fig:Tcwingab}(Color online) Magnetic field dependence of $d\rho/dT$ at the Curie temperature $T_{\textrm{C}}$. Full symbols indicate the evolution when the anomalies are positive. For $H>H_{\textrm{CP}}$, $d\rho/dT$ decreases smoothly with magnetic field, as a crossover of the PM-FM1 transition. Open symbols draw the evolution when the anomalies are negative. Lines are guide for the eyes. Lines are continuous in the case of the first order transition and dashed in the case of the crossover.}
\end{center}
\end{figure}

\begin{figure}[!bht]
\begin{center}
\includegraphics[width=8.5cm]{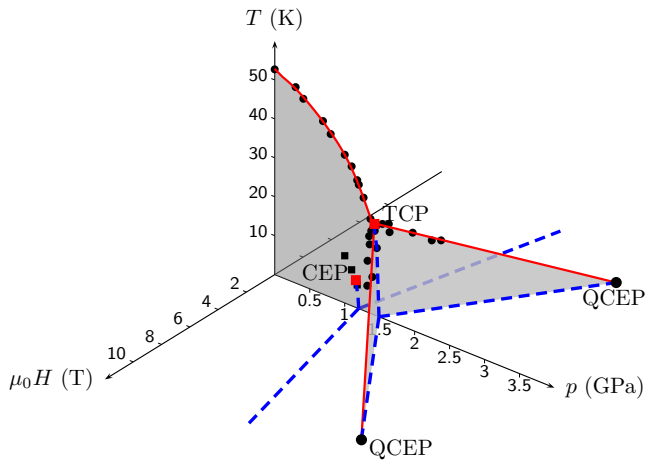}
\caption{\label{fig:diag3D}(Color online) Temperature-pressure-magnetic field phase diagram of UGe$_2$ drawn from resistivity measurements. Gray planes are planes of first order transition. Solid (red) lines are second order lines. First order ferromagnetic transition exists at finite field and temperature for $p>p_{\textrm{TCP}}$.}
\end{center}
\end{figure}

We conclude that there are two different anomalies with different behavior under the magnetic field. The positive peak is broadened and disappears under the magnetic field. The associated $T_{\textrm{C}}$ increases slowly under the magnetic field. In contrast, the minimum is more visible and associated $T_{\textrm{C}}$ increases rapidly. As we already mention, a PM-FM second order transition is changed to a crossover under magnetic field parallel to the magnetization. But as soon as the PM-FM transition is of first order, the magnetic field parallel to magnetization does not destroy the first order transition. These behaviors are presented for different pressures in Figs.~\ref{fig:drodTabcd}(d) and~\ref{fig:Tcwingab}, where full lines draw the evolution of the negative anomalies (minimum) and dashed lines follow the positive ones (peak). The boundary between the negative anomaly (open symbols) and the positive peak (full symbols) allows us to draw the second order transition line which limits the first order surfaces. Thus, the wings are experimentally plotted [see Fig.~\ref{fig:diag3D}], in good agreement with the schematic phase diagram~\cite{Uhlarz2004,Kimura2004,Pfleiderer2001,Belitz2005,Yamada2007,Rowley2010} for weak itinerant ferromagnets. Using a simple linear extrapolation, the second order transition occurs at $0$~K at a QCEP around $10-15$~T and $3-4$~GPa, but higher pressure measurements are required for more accurate determination.

The wing structure exists in any theory that describes a ferromagnetic first order transition~\cite{Belitz2005}. It was pointed out that a first order transition can occur if the Fermi level is between two peaks in the density of states~\cite{Shimizu1964}. This particular structure of the density of states can also provide an explanation for the FM1-FM2 transition~\cite{Sandeman2003}. Another theory shows that the effects of gapless particle-hole excitations at the Fermi surface induce a nonanalytic term in the Landau expansion of the free energy $F$ as a function of the magnetic moment $M$~\cite{Belitz1999}. It suggests that FM transitions in clean three-dimensional itinerant ferromagnets are always of first order at low enough temperature~\cite{Belitz1999}. The nonanalytic term comes from long-wavelength correlation effects and successfully explains the first order transition at low $T$ while the higher temperature transition is of second order~\cite{Belitz1999}. If the long range correlation effects are important, i.e., when the transition is first order, one may expect that these effects dominate the critical behavior measured by resistivity. With an itinerant model of the magnetic moment, long-range correlation effects can explain a negative anomaly for $d\rho/dT$~\cite{Su1975}. Then, at higher temperature, long-range effects do not dominate and the transition is of second order as for usual ferromagnets. A transition can also be of first order because of the magneto-elastic coupling~\cite{Bean1962,Yamada2007,Gehring2008}. It was shown that in a two-dimentional Ising lattice, the transition will become of first order if the exchange interaction is a function of lattice spacing and that the lattice is deformable~\cite{Rice1954}.

Indications of the FM wings have been reported recently from thermal expansion measurements realized via a strain gauge~\cite{Kabeya2010}. The detection via resistivity can be realized down to very low temperature while the use of strain gauge is limited to above $1.5$~K. As visible on Fig.~\ref{fig:diag3D}, the pressure extension of the PM-FM1 wings is very large by comparison to the zero field parameter~: $p_{\textrm{QCEP}}-p_{\textrm{TCP}}>>p_c-p_{\textrm{TCP}}$. This is due to the large values of the magnetization and of the magnetization jump at $p_c$ ($M_0=0.9\mu_B$ and $\Delta M_0=0.9\mu_B$) associated with large spontaneous magnetostriction ($10^{-4}/\mu_B$ in good agreement with the other heavy fermion compounds~\cite{Miyake2009}). This large pressure range makes UGe$_2$ a unique case to observe the FM wings. Such a phenomena is certainly difficult to observe in systems like UCoGe where the $M_0$ jump will be 1 order of magnitude smaller~\cite{Huy2007}.

In summary, we have shown that the wing structure phase diagram can be determined by resistivity measurements. The lines of critical points that link the TCP to the QCEP are located in the $p$-$T$-$H$ space. The next experimental challenge is to reach the QCEP at higher pressure and magnetic field.

We would like to thank L.~Malone, V.~Mineev, H.~Kotegawa and R.~Settai, and J.M. Martinod for useful discussions. This work was financially supported by the French ANR \mbox{CORMAT} and SINUS.


\end{document}